# Critical Capillary Number of Interfacial Film Displacement in a Capillary Tube

**Changfei Yan and Huihe Qiu***

Department of Mechanical and Aerospace Engineering

The Hong Kong University of Science & Technology,

Clear Water Bay, Kowloon, Hong Kong SAR, China

*Corresponding e-mail: meqiu@ust.hk

**Abstract**

The role of surface tension and wettability in the dynamics of air–liquid interfaces during immiscible fluid displacement flows in capillary tube driven by pressure has been investigated. The contact angle and capillary number drive the force wetting processes which is controlled by the balance between the capillary and the viscous lubrication forces. The dynamic wetting condition with the critical capillary number is studied analytically and validated experimentally, which demonstrates that the critical capillary number is associated with the contact angle, slip length and capillary radius.

*Keywords: Interfacial Film Displacement, Critical Capillary Number, Forced Dewetting, Contact Angle*

When a solid substrate is drawn from a liquid bath[1] in a low velocity (Forced Dewetting), the edge of the fluid (contact line) moves with the liquid due to gravity. However, when the velocity of the solid substrate is higher than a critical speed, a film of liquid remains on the substrate. Gravity is the inertial force (body force) in this fluid dynamics process. In micro channels, the viscous force and surface tension[2] force dominate fluid dynamics and gravity can be neglected. Normally, the Capillary number (viscous force/surface force) of a multi-phase flow in a macro/micro capillary is far less than one, which means surface force is dominant. A moving contact line is common in these small Capillary number physical processes[3]. When a gas bubble is pushing a liquid meniscus forward with a low capillary number in a capillary tube, a moving contact line will be expected as shown in Fig. 1 (a). In this case, no liquid film displacement will occur. However, a liquid film layer will be generated if the velocity of the meniscus (Fig. 1 b) is larger than a critical number. The phenomenon is quite similar to forced wetting in pulling substrate from a liquid bath, nevertheless, the dynamics force is surface force: pressure.

The Navier-Stokes equation with standard non-slip boundary conditions will produce an infinite energy dissipation close to the contact line[4][5][6]. The slip model of a moving contact line is widely accepted after C. Huh and L. Scriven[4] modeled a flat fluid steadily moving with creeping flow approximation. It is well known that the fourth-order nonlinear diffusion equation $\frac{\partial h}{\partial t} + h^n \frac{\partial^3 h}{\partial x^3} = 0$ is widely applied in interfacial film dynamics [7] ($h$ is the film thickness). Another equation[8] $\frac{Ca}{h^2+3\lambda h} = \frac{\partial^3 h}{\partial x^3} - \frac{\partial h}{\partial x} + 1$ using lubrication theory based on a slip boundary condition is taken as a classic result for the contact line on the moving plate from a liquid bath ($\lambda$ is the slip length). Several other theoretical researches about the moving contact line in a capillary driven by gravity (body force) were investigated[9][10]. Hocking[11] derived a critical capillary number prediction equation based on a body force driven model. To the authors' knowledge, no theoretical equation has as yet been derived for estimating the critical capillary number based on pressure driven conditions.

In this letter, we present the relationship of the critical capillary number and equilibrium contact angle $\theta_e$ [12] in a capillary when a liquid meniscus is driven by pressure. Although D Quere[13] presented a relationship in a capillary with an empirical equation, there is no solid theoretical support. We first deviated a hydrodynamic equation with the pressure $P_x$ based on a lubrication equation regarding the region around the contact line with the assumption that the equilibrium contact angle of the liquid film is small[23]. The equation is solved by matching the curvatures of inner region and outer region. Utilizing a novel laser interference method[14] for measuring the interfacial film thickness, experiments were conducted for validating the results calculated by the new equation.

Considering a pressure driven meniscus moving forward with speed $U$ in a capillary $u$ and $v$ are the velocity of liquid film in x and y directions, as shown in Fig. 1. The conservation of mass should be: $u_x + v_y = 0$, where the subscripts x and y denote the differentiation with respect to that variable. The Navier-Stokes equation in liquid film could be simplified dramatically by lubrication theory. The simplification limits the hydrodynamic equation deviated here to the case of a small equilibrium contact angle, which means a hydrophobic solid substrate is out of our consideration. The boundary conditions of the meniscus [4] in the capillary are:

$\mu u_y\big|_{y=h} = 0$ and $\lambda u_y\big|_{y=0} = u - U$, where $\lambda$ is the slip length and $U$ is the bulk velocity of the liquid meniscus. The boundary condition $\mu u_y\big|_{y=h} = 0$ is the assumption that the surface stress on the interface is equal to 0. After Greenspan[15] presented the small amount of slip in the vicinity of the contact line, Voniov[9] introduced the inner region's size as several times the size of a molecule. Cox[16] imposed the characteristic length of the inner region by slip length. The conservation of mass equation when averaged vertically across the thickness $h$ of the liquid film is:

$$\frac{\partial h}{\partial t} + \frac{\partial h \bar{U}}{\partial x} = \frac{\partial h}{\partial t} + U\frac{\partial h}{\partial x} - \frac{\partial}{\partial x}\left(\frac{P_x h^2}{3\mu}(h + 3\lambda)\right) = 0 \qquad (1)$$

$P_x$ is the pressure drop along the x direction.

When the behavior of liquid flow in the capillary is laminar, the velocity distribution of the fluid is parabolic. If the interface of the liquid meniscus does not change with time, the velocity distribution of the whole liquid could be treated as parabolic. This means that when $\frac{\partial h}{\partial t} = 0$, $P_x$ in equation (1) equals the pressure drop in a laminar continuous flow.

Considering the air flow faraway from the meniscus, one can assume there is a continuous constant flow in the capillary. $u$ is horizontal constant velocity, $v$ is vertical velocity, $v = 0$. The average velocity $\bar{U} = \int_0^r u dy$ is a constant value. According to the N-S equation, when $u, v = 0$ is constant, with the boundary conditions of a no-slip situation: $u\big|_{y=0, y=2r} = 0$ and $u\big|_{y=r} = u_{max}$, the average velocity is:

$$\bar{U} = \frac{1}{r}\int_0^r u dy = -\frac{r^2 P_x}{3\mu} \tag{2}$$

Because the flow in the capillary is laminar and continuous, $P_x$ is constant. The value should be:

$P_x = -\frac{3\gamma k Ca}{r^2}$. Where $Ca = \frac{\mu u}{\gamma}$ is the capillary number, $\mu$ is the dynamic viscosity of liquid, $u$ is velocity, $\gamma$ is the surface of interfacial tension between gas and liquid, $k = \frac{\mu_{air}}{\mu_{liquid}}$ is the ratio of dynamics viscosity. Neglecting the compressibility of the air flow, we can take the average velocity $\bar{U}$ as the velocity of the meniscus $U$. When $h_x \ll 1$, the pressure drop in the liquid near the meniscus is caused by the pressure drop in the air and the surface tension of the interface curve. Then equation (1) becomes:

$$3\mu \frac{\partial h}{\partial t} + 3\mu U \frac{\partial h}{\partial x} - \frac{\partial}{\partial x}\left(\left(\gamma h_{xxx} + \frac{3\gamma k Ca}{r^2}\right)h^2(h + 3\lambda)\right) = 0 \tag{3}$$

The quantities appearing in equation (3) can be non-dimensionalized in terms of the capillary inner radius $r$ as follows:

$$h = rh_r,\ x = rx_r,\ \lambda = r\lambda_r,\ t = rt_r/U \tag{4}$$

If the contact line is steady in the laboratory frame[11], equation (3) is time independent ($\frac{\partial h_r}{\partial t_r} = 0$), and then equation (3) becomes:

$$\frac{3Ca}{h_r^2 + 3\lambda_r h_r} = h_{r_{x_r x_r x_r}} + 3kCa \tag{5}$$

Equation (5) is an equation concerning the liquid profile deviated from lubrication theory and thin film assumption. So, it is an equation describing the inner region around the contact line. The curvature of inner region and outer region should be the same at the boundary[17]. If we take equation (5) as a normal high-order differential equation, the condition coming from outer region could be treated as a boundary condition. As the liquid meniscus is moving in a capillary tube, the maximum curve of the interface is $\frac{1}{r}$. Because the inner region's character length is slip length $\lambda$, we set the boundary condition of the outer region at $3\lambda$: $h_{xx}|_{x=3\lambda} = \frac{1}{r}$. Although the exact analytic solution for equation (5) is not possible to achieve, we can get the approximate solution by simplification and perturbation theory.

On the one hand, in the inner region near the contact line, where $h$ is small and the interface is highly curved, the left hand side of equation (5) is balanced by the first term on the right: viscous forces are balanced by surface tension forces. Slip length[18] is a function of shear rate. Although the exact slip length value in normal wettability material is still unknown, the scale of slip length could be set as 2nm, comparing with the scale of slip length of super hydrophobic material is 20nm by Craig *et.al.*[19]. Bretherton[20] presented the relationship between the capillary number and liquid film thickness in a capillary. When the capillary number is less than 0.01, the scale of film thickness is $h \sim 1\mu m$, according to his presentation. In the inner region, $h^2 \gg 3\lambda h, \frac{3Ca}{h^2+3\lambda h} \gg 3kCa$. The equation near the contact line reduces to

$$\frac{3Ca}{h_r^2} = h_{r_{x_r x_r x_r}} \qquad (6)$$

Let $\varepsilon = 3Ca$, $h_r = \varepsilon^{\frac{1}{3}} y$, according to Duffy & Wilson[21], the general solution to the third-order ordinary differential equation (6) is:

$$y_{x_r x_r} = \left(\frac{\beta 2^{1/6}}{\pi Ai(s_1)}\right)^2 \qquad (7)$$

Where $\beta$ is a constant number, $Ai$ is the Airy function, $s_1$ is a root of equation $\alpha Ai(s_1) + \beta Bi(s_1) = 0$. The first-order differential equation $h_{r_{x_r}}$ is a function of $x_r$:

$$h_{r_{x_r}}^3 = \varepsilon y_{x_r}^3 \approx 3\varepsilon \ln\left(\frac{\pi}{2^{2/3} x_r \beta^2}\right) = 9Ca \ln\left(\frac{\pi}{2^{2/3} x_r \beta^2}\right) \qquad (8)$$

On the other hand, we can simplify equation (5) by replacing $\lambda_r$. let

$$h_r = 3\lambda_r h_{r\lambda}, \; x_r = 3\lambda_r x_{r\lambda} \tag{9}$$

then equation (5) becomes:

$$\frac{3Ca}{h_{r\lambda}^2 + h_{r\lambda}} = h_{r\lambda_{x_{r\lambda}x_{r\lambda}x_{r\lambda}}} + 3Cak(3\lambda_r)^2 \tag{10}$$

When $P_x = 0$, contact angle is determined by Young's equation according to thermodynamics. $h_{r\lambda_{x_{r\lambda}}}$ equates to the equilibrium contact angle at origin [22]. The boundary condition of the perturbation expansion of equation (10) at the contact line when $Ca = 0$ should be $h_{r\lambda}|_{x=0} = 0$, $h_{r\lambda_{x_{r\lambda}}}|_{x=0} = \theta_e$, $h_{r\lambda_{x_{r\lambda}x_{r\lambda}}}|_{x=0} = 0$. Then, we have

$$h_{r\lambda_{x_{r\lambda}}}^3 = \theta_e^3 + 3\theta_e\varepsilon\left(\frac{x_{r\lambda}lnx_{r\lambda}}{\theta_e} - \frac{\theta_e x_{r\lambda}+1}{\theta_e^2}ln(\theta_e x_{r\lambda}+1) - \frac{9k\lambda_r^2 x_{r\lambda}^2}{2} - c_1 x_{r\lambda} - c_2\right) + O(\varepsilon^2)$$

$$\tag{11}$$

Comparing with equation (8) and equation (11), we find:

$$h_{r\lambda_{x_{r\lambda}}}^3 = \theta_e^3 + 3\theta_e\varepsilon\left(\frac{x_{r\lambda}lnx_{r\lambda}}{\theta_e} - \frac{\theta_e x_{r\lambda}+1}{\theta_e^2}ln(\theta_e x_{r\lambda}+1) - \frac{9k\lambda_r^2 x_{r\lambda}^2}{2} - c_1 x_{r\lambda} - c_2\right) = h_{r_{x_r}}^3 =$$

$$\varepsilon y_{x_r}^3 \approx 3\varepsilon \ln\left(\frac{\pi}{2^{2/3} x_r \beta^2}\right) \tag{12}$$

As the boundary condition of the outer region of the approximate solution (12) is $h_{xx}|_{x_{r\lambda}\sim 1} = \frac{1}{r}$, in short:

$$exp(\frac{\theta_e^3}{9Ca} - \left(1+\frac{1}{\theta_e}\right)ln(\theta_e+1) - \frac{9k\lambda^2}{2\theta_e r^2} - \frac{c_1}{\theta_e} - c_2) = \frac{(3Ca/2)^{1/3}r}{3\lambda\pi Ai^2} \tag{13}$$

In equation (13), $c_1$ and $c_2$ are constant number from integral calculation. Normally, these constant numbers can be determined by the boundary condition in integral method. However, the boundary condition in the perturbation theory equation (10) is the boundary condition for $h_{r\lambda_0}$. For determining $c_1$ and $c_2$, we consider adding in two other boundary conditions for the small capillary number:

$$(a)\ h_{r\lambda_{x_{r\lambda}}}(0) = \theta_e,\ (b)\ h_{r\lambda_{x_{r\lambda}x_{r\lambda}}}(x_{r\lambda} \sim 1) = 0 \tag{14}$$

In physics, equation (14a) means the slope of the thin liquid film at the contact line will not change when the capillary number increases from 0 to a small value; equation (14b) means the curvature of the thin liquid film is constant in the inner region from the contact line to outer region. As equation (7) points out that the slope near the contact line varies linearly with x, the boundary condition equation (13) does make sense. With these two boundary conditions, it's easy to find: $c_1 = \frac{1}{\theta_e} \ln\left(\frac{\theta_e}{\theta_e+1}\right), c_2 = 0$. Thus, equation (13) becomes:

$$\exp\left(\frac{\theta_e^3}{9Ca} - \left(1 + \frac{1}{\theta_e}\right)\ln(\theta_e + 1) - \frac{9k\lambda^2}{2\theta_e r^2} - \frac{1}{\theta_e^2}\ln\left(\frac{\theta_e}{\theta_e+1}\right)\right) = \frac{(3Ca/2)^{1/3}r}{3\lambda\pi Ai^2} \tag{15}$$

Equation (15) is a function of slip length, the critical capillary number and equilibrium contact angle. To verify that the formula is correct, experiments are performed with different capillary numbers in capillaries with different equilibrium contact angles.

In the experiments, after filling the capillary tube with working fluid (deionized water and ethanol), a long air bubble is injected into the center of the capillary. After leaving it to stand for a while, a meniscus of working fluid is generated at the head of the air bubble. By monitoring the liquid film near the contact line under different capillary numbers, the critical capillary numbers for different equilibrium contact angles can be measured.

A recently developed laser interference measurement technique [14] is used for measuring liquid film thickness behind the contact line when the meniscus is pushing forward by pressure. As shown in Fig. 2, the experiment set up contains an optical measurement system, a pressure support part, an acrylic triangle tank full of glycerol and a high speed camera. The laser in the optical measurement system is the ILT Model 5500A Air-cooled Argon Ion Laser. To filter other unexpected wavelength lasers, a polarized beam splitter is applied to block 488nm laser in the optical measuring system. After collimating, the laser beam (514nm) is extended into a laser sheet horizontally by two cylinder lenses. The third cylinder lens between the glycerol triangle tank and the high speed camera aims to make the parallel laser beams coherent with each other on the camera's CCD. A Harvard pump 33 Dual syringe pump and a syringe are used for the pressure support part. The glycerol in acrylic right triangle tank (refractive index: 1.474) is made to match the refractive index of capillary glass. The recording rate of the high speed camera (Redlake MotionXtra HG-100 K) is set to 1000fps with a resolution of 1504×1128 pixels. The refractive indices of the deionized water and air at 514nm are $n_2 = 1.33, n_3 = 1$, respectively. The angle between the camera's normal and laser beam is set as 120° to maximize the measured resolution.

Three different materials are used for the capillary walls in the experiment: borosilicate glass (B100-50-10), natural hard glass and FAS (fluoroalkylsilane) coated borosilicate glass. The inner diameter of the three capillaries in all the experiments is 0.5mm. The equilibrium contact angles of the pure deionized water on natural hard glass, borosilicate glass and FAS coated borosilicate glass capillary are 13.5°, 26°, and 110° respectively. The range of capillary numbers of the bulk fluid meniscus is set from 0.0001 to 0.03. The mixture of ethanol and deionized water in this letter is only tested in the capillary coated with FAS. The contact angle of the mixtures on the FAS capillary is shown in Table 1.

The experimental results are shown in Fig. 3. "X% FAS" denotes that the experimental results were measured when the x% concentration ethanol mixture was moving in an FAS coated capillary. When the capillary number goes down, a swoop exists. This is the critical capillary number for the different contact angles.

The critical capillary numbers of mixtures of water and ethanol in different capillaries are shown in Fig 4. In equation (15), $\lambda \sim 2$nm, $r = 0.25mm$, we assume $Ai$ as its global maximum 0.53566[23]. Compared to the theoretical capillary number, this experiment result matches the equation (15) very well. Due to the uncertainty of the laser interference measurement technique being 0.3 μm, the liquid film of pure water in natural hard glass is the minimum value measured in experiment. The experiment critical capillary number when the contact angle equates to 110° is 0.002. As discussed above, equation (15) is only validated on a hydrophilic substrate. The critical capillary number of deionized water in the FAS coated capillary calculated is only for reference.

D. Quere[13] and Hocking[11] predicted the critical capillary number is a linear function of cubic contact angle. Compared with the right hand part of Hocking's equation (10), equation (15) displaces the gravity force to surface force, which means the second term in the left hand side of equation (15) cannot be neglected when the capillary number tends to zero as with Hocking's, because the surface force will dominate. This term causes the curve to be lower than Hocking's as shown in Fig 4. The contact angle in Hocking's paper is the equilibrium contact angle, which is different to D. Quere's receding contact angle. D. Quere's empirical equation (blue line in Fig 4) is much lower than the experiment data here. Surface roughness is a reason for D. Quere's estimates. Another explanation for D. Quere's equation is that the partial wettability in capillary is not uniform. Once some tiny areas' contact angle is smaller than the average number, the slug will shrink when passing these areas. However, D. Quere measurement method (checking the change in length of the slug) cannot overcome this problem. Equation (15) also shows that the capillary radius is another important parameter in prediction.

In conclusion, for the first time, we derived the theory regarding the critical capillary number with the wettability effect in a pressure driven capillary tube. A lubrication model is utilized in the theoretical deviation, which limits the equations here on a hydrophilic surface. By matching the curvatures of the inner and outer regions, we can obtain an equation for predicting the critical capillary number. The critical capillary numbers calculated by this new equation match with the experiment data very well.

This work was supported by the Research Grants Council (RGC) of the Hong Kong Special Administration Region, China (Grant Nos. 617812 and 16207515).

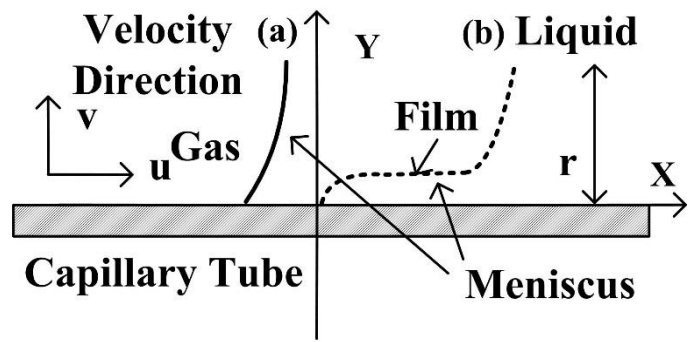

Figure 1 Schematic of meniscus moving in a capillary: (a) with low capillary number; (b) with high capillary number

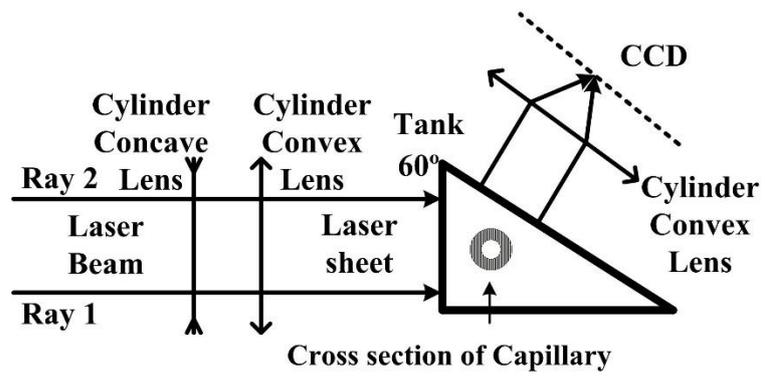

Figure 2. The schematic for interference laser measure method

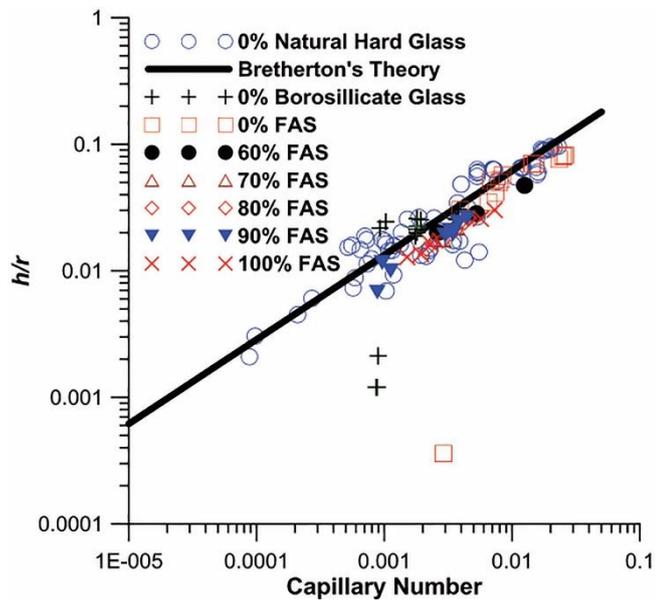

Figure 3. The experiment measured liquid film thickness (film thickness/radius) and capillary number

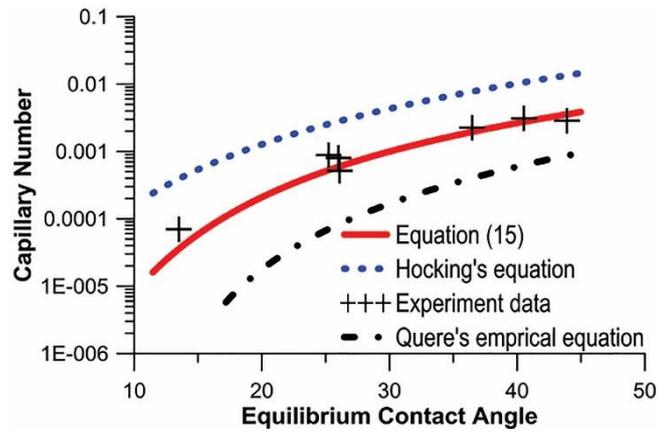

Figure 4. The relationship of critical capillary number and equilibrium contact angle

Table 1. Properties of ethanol and deionized water mixture

| Ethanol Concentration | 100% | 90% | 80% | 70% | 60% |
|---|---|---|---|---|---|
| Surface tension ($10^{-3}$ N/m) | 22.27 | 23.49 | 24.87 | 26.31 | 27.76 |
| Viscosity ($10^{-3}$ Pa·s) | 1.14 | 1.65 | 2.18 | 2.58 | 2.79 |
| Contact angle (±1°) | 26.09 | 25.24 | 36.47 | 40.52 | 43.91 |

References:

[1] D. Bonn, J. Eggers, J. Indekeu, J. Meunier and E. Rolley, Rev. Mod. Phys. 81, 739 (2009).
[2] T. Myers, SIAM Rev. 40, 441 (1998)
[3] Y. Sui, H. Ding and P. Spelt, Annu. Rev. Fluid Mech. 46, 97(2014).
[4] C. Huh and L. Scriven, Journal Of Colloid And Interface Science 35, 85 (1971).
[5] R. Pit, H. Hervet, and L. Le´ger, Phys. Rev. Lett. 85, 980 (2000).
[6] A. Alizadeh Pahlavan, L. Cueto-Felgueroso, G. H. McKinley and R. Juanes, Phys. Rev. Lett. 115, 034502 (2015).
[7] N. SMYTH and J. HILL, IMA Journal Of Applied Mathematics 40, 73 (1988).
[8] L. HOCKING, The Quarterly Journal Of Mechanics And Applied Mathematics 36, 55 (1983).
[9] O. Voinov, Fluid Dynamics 11, 714 (1976).
[10] R. Sedev and J. Petrov, Colloids And Surfaces 53, 147 (1991).
[11] L. HOCKING, European Journal Of Applied Mathematics 12, 195 (2001).
[12] P.G. de Gennes, Rev. Mod. Phys. 57, 827 (1985).
[13] D. Que´re´, C. R. Acad. Sci. Paris, Se´rrie II 313, 313 (1991).
[14] C. Yan and H. Qiu, Measurement Science And Technology 27, 065204 (2016).
[15] H. Greenspan, J. Fluid Mech. 84, 125 (1978).
[16] R. Cox, J. Fluid Mech. 168, 169 (1986).
[17] J. Eggers, Physics Of Fluids 16, 3491 (2004).
[18] N. V Priezjev, Physical Review E 80, 031608 (2009).
[19] V.S.J. Craig, C. Neto and D.R.M. Williams, Phys. Rev. Lett. 87, 054504 (2001).
[20] F. Bretherton, J. Fluid Mech. 10, 166 (1961).
[21] B. Duffy and S. Wilson, Applied Mathematics Letters 10, 63 (1997).
[22] R. Tadmor, Langmuir 20, 7659 (2004).
[23] E.Lorenceau, D.Quere, and J.Eggers, Phys. Rev. Lett. 93, 254501 (2004).